\documentclass[aps,prx,showpacs,twocolumn,reprint,superscriptaddress]{revtex4-2}

\usepackage{qcircuit}
\usepackage[dvips]{graphicx}
\usepackage{amsmath,amssymb,amsthm,mathrsfs,amsfonts,dsfont,mathtools}
\usepackage{epsfig}
\usepackage{braket}
\usepackage{hyperref}
\usepackage{bm}
\usepackage{enumerate}
\usepackage{color}
\usepackage{graphicx}
\usepackage{pgf}
\usepackage{pgfplots}
\pgfplotsset{compat=1.18}
\usepackage{tikz}
\usepackage{enumitem}
\usepackage[capitalize]{cleveref}
\usepackage[normalem]{ulem}

\newcommand{\sign}{\text{sign}}
\newcommand{\tr}{\mathrm{Tr}}

\newcommand{\var}{\mathrm{Var}}

\newcommand{\ucirc}{\mathcal{U}_{circ}}
\newcommand{\ucirce}{\hat{\mathcal{U}}_{circ}}
\newcommand{\nshot}{N_s}

\usepackage{amsthm}

\newtheorem{statement}{Statement}
\newtheorem{theorem}{Theorem}

\Crefname{theorem}{Theorem}{Theorems}
\theoremstyle{remark}

\newcommand{\qmaddress}{Quantum Motion, 9 Sterling Way, London N7 9HJ, United Kingdom}

\begin{document}	
	
\title{Probabilistic Interpolation of Quantum Rotation Angles}

\author{B\'alint Koczor}
\email{balint.koczor@materials.ox.ac.uk}
\affiliation{\qmaddress}
\affiliation{Department of Materials, University of Oxford, Parks Road, Oxford OX1 3PH, United Kingdom}
\author{John Morton}
\affiliation{\qmaddress}
\affiliation{London Centre for Nanotechnology, UCL, 17-19 Gordon St, London WC1H 0AH, United Kingdom}
\author{Simon Benjamin}
\email{simon.benjamin@materials.ox.ac.uk}
\affiliation{\qmaddress}
\affiliation{Department of Materials, University of Oxford, Parks Road, Oxford OX1 3PH, United Kingdom}

\begin{abstract}
Quantum computing requires a universal set of gate operations; regarding gates as rotations, any rotation angle must be possible. However a real device may only be capable of $B$ bits of resolution, i.e. it might support only $2^B$ possible variants of a given physical gate. Naive discretization of an algorithm's gates to the nearest available options causes coherent errors, while decomposing an impermissible gate into several allowed operations increases circuit depth. Conversely, demanding higher $B$ can greatly complexify hardware. Here we explore an alternative: Probabilistic Angle Interpolation (PAI). This effectively implements any desired, continuously parametrised rotation by randomly choosing one of three discretised gate settings and postprocessing individual circuit outputs. The approach is particularly relevant for near-term applications where one would in any case average over many runs of circuit executions to estimate expected values. While PAI increases that sampling cost, we prove that a) the approach is optimal in the sense that PAI achieves the least possible overhead and c) the overhead is remarkably modest even with thousands of parametrised gates and only $7$ bits of resolution available. This is a profound relaxation of engineering requirements for first generation quantum computers where even $5-6$ bits of resolution may suffice and, as we demonstrate, the approach is many orders of magnitude more efficient than prior techniques. Moreover we conclude that, even for more mature late-NISQ hardware, no more than $9$ bits will be necessary.
\end{abstract}

\maketitle

\section{Introduction}

Producing quantum computers of the scale and fidelity needed to solve practically useful problems requires development not just of the quantum processor itself, but of the analogue and digital electronics used for the control and readout of qubits. These electronics may include field-programmable gate array (FPGA) systems~\cite{fpgacontrol1,fpgacontrol2} and customised integrated circuits (ICs)~\cite{Das2011LowPF,cryoCMOSvandijk, cryoCMOSbardin, cryoCMOSreilly}, typically cooled to improve performance and integration with the qubits.
Understanding and minimising the required specifications of the electronics supporting the quantum computer is an essential step in developing scalable systems in general; These issues, however, become even more acute when considering cryogenic electronics with limited power budgets~\cite{vandijk1}, and/or quantum processor architectures targeting close integration of quantum systems with classical systems, such as for silicon-based spin qubits~\cite{CMOSreview}.

Consider, for example, the instruction to implement a parametrised Pauli gate in which a user has specified the kind of gate they want to implement and to which qubits.
Parametrised Pauli gates are naturally implemented in most physical platforms and the ability to realise a continuous set of parametrised gates is required by most near-term quantum algorithms~\cite{cerezoVariationalQuantumAlgorithms2021a, endoHybridQuantumClassicalAlgorithms2021, bharti2021noisy}. 
As the gate angles are defined and implemented using digital electronics, the gate angles must be discretised into $B$ bits of resolution.
The choice of $B$ has significant impacts on elements such as the bandwidth of communication channels between different elements of the control stack; the memory requirements of any gate instruction cache; and in the digital-to-analogue converters (DACs) used ultimately to produce the driving fields acting on the qubits. 
There is therefore a strong benefit in minimising $B$ to the point where it is just sufficient to provide the required gate fidelities for a given application or circuit. 
Most of the currently leading qubit hardware platforms operate optimally at cryogenic temperatures, including superconducting qubits~\cite{cryoCMOSbardin, supercqubits}, trapped ions~\cite{cryooinons,trappedion2}, semiconductor spin qubits~\cite{cryosi, sixqubitsi}, and photonic qubits~\cite{cryophotonics}. This has motivated significant effort on developing control systems which can also operate adjacent to the qubits, at low temperatures, where the motivation to minimise power consumption becomes even greater.  

Previous studies have examined the required hardware specifications for qubit control, with the goal of optimising the fidelity of individual logic gates to the point that they are limited by factors other than (e.g.) the bit resolution of the control system~\cite{vandijk1}. However, focusing on the impact of control limitations on a single gate operation, rather than on the average output of a quantum calculation, risks significantly over-specifying the control hardware requirements.
Our approach is to consider the output state of the quantum device, and examine its sensitivity to the number of bits used in the angular discretisation of the gates (which can ultimately be related to parameters such as the bit resolution and/or sampling rate of a qubit control DAC).

In principle even $3$ bits of angular resolution 
would already guarantee a universal computing machine, as any
continuously parameterised gate can be arbitrarily well approximated
as a sequence of discrete Clifford and T gates \cite{dawson2005solovay}. However, realising the desired gate by a sequence of discrete options, one would significantly deepen the overall circuit; this is undesirable in general and particularly so for near-term quantum 
computers where one always prefers shallow circuits. Using a method of Probabilistic Angle Interpolation (PAI) we show that we can effectively upgrade the capabilities of a physical device with only a set of discrete angles such
that on average it can implement any continuous rotation gate -- and this is achieved without increasing circuit depths. We do so by randomly instructing
the control infrastructure to perform one of the discretised rotation angles, and we subsequently
combine the individual outputs of the device
such that on average we obtain the same output as the ideal device with infinite resolution. 
The PAI approach thus allows one to fully exploit the power of algorithms that (nominally) require continuously parametrised gates.
It does so with only a marginal increase in repetition (sampling) cost for any reasonable number of parameters as long as the
aim is to estimate expected values of observables as relevant in most near-term quantum algorithms~\cite{cerezoVariationalQuantumAlgorithms2021a, endoHybridQuantumClassicalAlgorithms2021, bharti2021noisy}.

Note that our PAI technique addresses the challenge of discretisation, rather than the well-studied issue of gate infidelity -- i.e. random and unknown variations in the implemented gate due to, e.g.,
imperfections or interactions with the environment.
For the latter case techniques such as probabilistic error cancellation (PEC), are commonly used in the context of
quantum error mitigation~\cite{cai2022quantum} whereby noise in a quantum gate is mitigated by learning the
noise model of the particular gate or set of gates and randomly inserting recovery operations such that their average effect
cancels out the effect of noise -- we discuss further connections in \cref{sec:discussion}.
In the following analysis, however, we assume perfect unitary gates and thus
the only challenge we address is that the rotation angles of quantum gates are discretised for the sake
of reducing the complexity
of engineering. Needless to say, methods of error mitigation can be combined with our PAI techniques. 

\section{Probabilistic Angle Interpolation}
\subsection{Summary of the protocol\label{sec:summary}}
Focusing on quantum systems of $N$ qubits,
we consider parametrised quantum gates $R(\theta)=e^{-i \theta G/2}$ with 
gate generators $G$ of eigenvalues $\pm1$; These include parametrised SWAP gates
and Pauli gates for any Pauli string $G \in \{ \openone, X, Y, Z \}^{\otimes N}$.
These gates encompass most gatesets developed for quantum technologies, such as
single qubit $X$, $Y$ or $Z$ rotations or two-qubit $XX$ entangling gates. 
As we presently explain, the PAI method can also be extended to other physical gate sets.
We denote the superoperator of our parametrised gates as $\mathcal{R}(\theta)$ 
which acts by conjugation as $\mathcal{R}(\theta) \rho :=  e^{-i \theta P_k/2} \rho e^{i \theta P_k/2}$.

\begin{figure}[tb]
	\begin{centering}
		\includegraphics[width=0.4\textwidth]{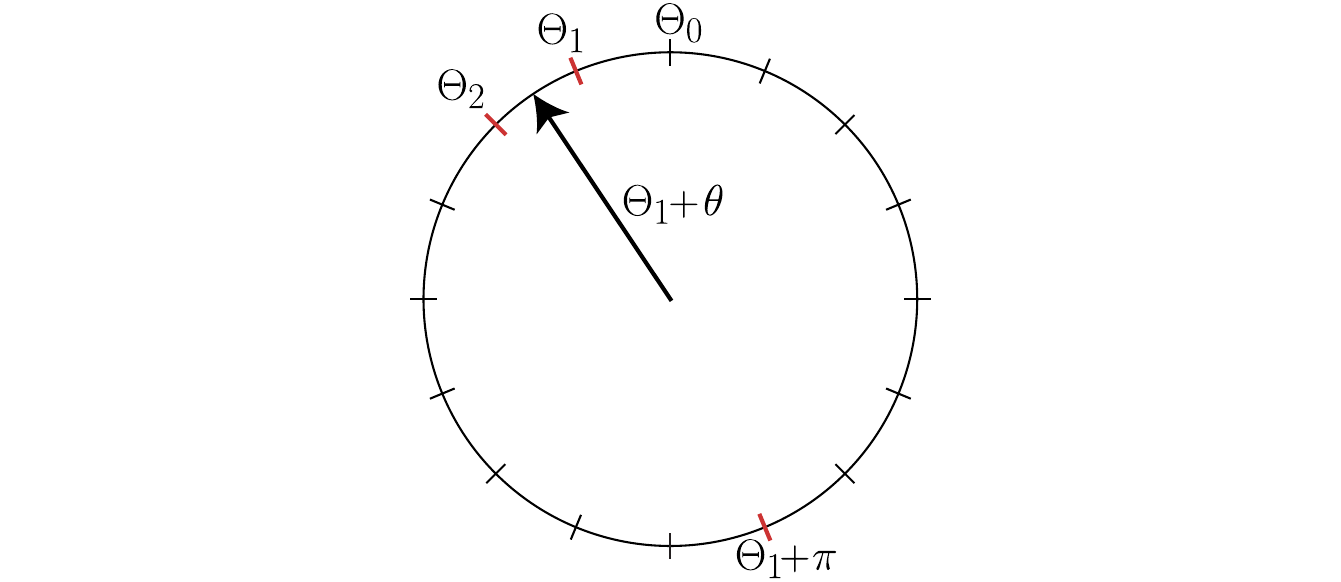}
	\end{centering}
	\caption{ 
		Continuously parametrised Pauli rotation gates $\mathcal{R}(\theta)$ encompass most typical
		gates developed for quantum technologies. Due to the use of digital electronics,
		the rotation angles are divided into $2^B$ equiangular segments $\Theta_k$. In order to
		reduce engineering complexity, the number $B$ of bits is chosen as small as possible.
		PAI realises an arbitrary, continuous rotation $\mathcal{R}(\Theta_k {+} \theta)$
		by randomly instructing the quantum hardware to apply one of the two nearest notch settings 
		$\mathcal{R}(\Theta_k)$ and $\mathcal{R}(\Theta_{k+1}) $ or with a small probability to 
		apply the antipolar rotation $\mathcal{R}(\Theta_k{+}\pi) $.
		\label{fig:fig1}
	}
\end{figure}

\cref{fig:fig1} illustrates a physical device that can perfectly perform
parametrised gates $\mathcal{R}(\Theta_k)$
but only with a finite angular resolution of $B$ bits as
\begin{equation}\label{eq:delta}
\Theta_k =  k \Delta \quad \text{with} \quad  \Delta = \frac{2\pi}{2^B}, \quad k \in \{ 0, 1, \dots  2^B-1  \}.
\end{equation}
and we detail below the generalisation to non-uniformly distributed (non-linear) set of angles.  
We define any continuous rotation angle 
as an overrotation of one of the discrete settings
$\mathcal{R}(\Theta_k {+} \theta)$  by an angle $0 \leq \theta  < \Delta$.
Given a relative position $\lambda = \theta/\Delta$ between two discrete settings, the most simple solution would be to round to the nearest notch; however this leads to systematic coherent errors, and in the numerics that we later present we observe that this effect can be remarkably severe.
A less naive approach would involve randomly switching between
two nearest notch settings $\Theta_k$ and $\Theta_{k+1}$ with probabilities
 $\lambda$ and $(1-\lambda)$, respectively, see \cref{app:approximate}.
However, on average one obtains a non-unitary operation that can result
in an exponential decrease of the fidelity of the quantum state
with the number of parametrised gates $\nu$ in the circuit. 

PAI randomly chooses one of three allowed
notch settings for each parametrised gate in a circuit
and exactly implements the desired continuous rotation angle by post processing measurement outcomes.
In particular, in each circuit execution we randomly choose either the nearest two notch settings
$\Theta_k$ and $\Theta_{k+1}$ or with a small probability we choose
the antipolar angle setting $\Theta_k{+} \pi$ as illustrated in \cref{fig:fig1}.
When estimating expected values with PAI, the individual circuit outputs are multiplied by a sign $-1$ 
whenever the third rotation angle was chosen.

Thus, the expected value estimation yields a probability distribution in \cref{fig:numerics} (left, grey histogram)
that is centred around the same mean value 
that one would obtain via an infinite angular resolution (blue).
However, the ability to exactly implement continuous rotations while having access to only
discrete rotation angles comes at the price of an increased 
number of circuit repetitions that scales in the worst case exponentially as $e^{\nu \Delta^2 /4}$
with the number of gates $\nu$.
We find in \cref{fig:overhead}, however, that at $B=7$ bits of resolution this overhead
is still reasonable when the number of parametrised gates in the circuit is not more than a few thousand
(of course, there can be arbitrarily many additional non-parameterised gates which align to the naturally available rotations).

\subsection{PAI of a single rotation gate \label{sec:single_gate}}

Introducing the notation for the aforementioned discrete notch settings as
\begin{equation}\label{eq:rots}
	\mathcal{R}_1 := \mathcal{R}(\Theta_k), \, \, \mathcal{R}_2 := \mathcal{R}(\Theta_{k+1}), \,\,
	\mathcal{R}_3 := \mathcal{R}(\Theta_k {+} \pi),
\end{equation}
the main observation we build on is that we can exactly express any overrotation $\mathcal{R}(\Theta_k {+} \theta)$
as a linear combination of the discrete gates as
\begin{equation}\label{eq:quasiprob}
	\mathcal{R}(\Theta_k {+} \theta) = \gamma_1(\theta) \mathcal{R}_1 + \gamma_2(\theta) \mathcal{R}_2
	+  \gamma_3(\theta) \mathcal{R}_3.
\end{equation}
By solving a system of trigonometric equations, we obtain the analytic form of the coefficients
$\gamma_l(\theta)$ in \cref{eq:solution} as a function of the continuous angle $\theta$. 
In a fashion analogous to quasiprobability sampling methods ~\cite{cai2022quantum,gambetta_error_mitig,practical_QEM} which mitigate non-unitary error effects,
our angular decomposition leads to the following implementation.
\begin{statement}\label{stat:one_gate}
We define a sampling scheme whereby we randomly choose one of the three discrete gate variants $\{\mathcal{R}_l\}_{l=1}^3$ 
from \cref{eq:rots} according to the probabilities $p_l(\theta) = |\gamma_l(\theta)| / \lVert \gamma(\theta) \rVert_1$
which yields the unbiased estimator of the rotation gate as
\begin{equation}\label{eq:estimator}
	\hat{\mathcal{R}}(\Theta_k {+} \theta) =  \lVert \gamma(\theta) \rVert_1  \, \mathrm{sign}[ \gamma_l(\theta) ] \mathcal{R}_l,
\end{equation}
such that $\mathbb{E}[\hat{\mathcal{R}}(\Theta_k {+} \theta)] = \mathcal{R}(\Theta_k {+} \theta)$.
\end{statement}
To intuitively understand the above approach, we expand the trigonometric probabilities to leading order in the small $\Delta$.
With probability $p_1(\theta) = (1- \lambda) + \mathcal{O}(\Delta^2)$ we apply the gate at the notch setting $\Theta_{k}$,
and with probability $p_2(\theta) = \lambda + \mathcal{O}(\Delta^2)$ we apply the gate at the next notch setting
$\Theta_{k+1}$. In leading order this would be equivalent to the naive approach discussed in the previous section which led
to a non-unitary operation.
In contrast, we obtain the desired, unitary operation by applying the antipolar rotation
$\mathcal{R}(\Theta_k {+} \pi)$ with a small probability $p_3(\theta) = \tfrac{1}{4} \lambda (1-\lambda) \Delta^2 + \mathcal{O}(\Delta^4)$.
Additionally, we multiply any observable measurement-outcome with the factor $\lVert \gamma(\theta) \rVert_1  \, \mathrm{sign}[ \gamma_l(\theta) ]$.

In \cref{theo1} we explicitly prove the above approach is optimal 
in the sense that it yields a minimal $\lVert \gamma \rVert_1$ and 
present a general solution that can be applied to non-uniform notch settings too.
	\begin{theorem}[informal summary of \cref{theo2}]\label{theo1}
		Given any set $\{ \mathcal{R}(\Theta_q) \}$ of discrete (possibly non-uniform)
		notch settings that a machine can realise,
		the optimal protocol that minimises $\lVert \gamma \rVert_1$ uses
		$\Theta_k$ and $\Theta_{k+1}$ as the two nearest notch settings to $\theta$
		and we choose the third gate to be the notch setting
		nearest to
		$\Theta_{k} {+} \pi {+} \frac{\Delta}{2} $,
		where we defined the distance $\Delta:= \Theta_{k+1} {-} \Theta_{k}$.
	\end{theorem}

\subsection{PAI of parametrised circuits}

We now consider a quantum circuit $\ucirc$ that contains $\nu$ continuously parametrised gates
and additionally may also contain other non-parameterised gates.
We apply \cref{stat:one_gate} to each of the continuously parametrised gates:
Given that each parametrised gate has a desired continuous rotation angle $\Theta_{k_l} {+}\theta_l$,
we first determine the corresponding notch settings $( \Theta_{k_1}, \Theta_{k_2} \dots \Theta_{k_\nu} )$
and corresponding overrotation angles $(\theta_1, \theta_2, \dots \theta_\nu)$ in \cref{eq:quasiprob}. 
At each execution of the circuit we randomly replace a parametrised gate with the
corresponding discrete gate variant, i.e, the $j^{th}$ parametrised gate is replaced 
by one of the  discrete gate variants $\mathcal{R}_{l_j}^{(j)}$ from \cref{eq:rots} according
to the probability distribution $p_{l_j}(\theta_j)$ from \cref{stat:one_gate}.

The result is a set of circuit variants
$\mathcal{U}_{\underline{j}}$ that contain only discrete notch settings
according to the multi index $\underline{j} = (j_1, j_2, \dots j_\nu) \in 3^\nu$.
\begin{statement} \label{stat:circ}
Given a circuit $\ucirc$ of $\nu$ parametrised gates
we choose a multi index $\underline{l} \in 3^\nu$ according to the probability
distribution  $p(\underline{l}) = |g_{\underline{l}}| / \lVert g  \rVert_1 $
where $g_{\underline{l}}$ are simply products of the single-gate
factors from \cref{stat:one_gate}.
We obtain an unbiased estimator of the ideal circuit as
\begin{equation}
	\ucirce 	= \lVert g  \rVert_1  \sign(g_{\underline{l}} ) \mathcal{U}_{\underline{l}},
\end{equation}
by executing the circuit variants $\mathcal{U}_{\underline{l}}$
in which all continuously parametrised gates are replaced by
the discrete ones according to the multi index $\underline{l}$.
Thus, $\mathbb{E}[\ucirce ] = \ucirc$.
\end{statement}
The above scheme can be compared to Probabilistic Error Cancellation (PEC)~\cite{gambetta_error_mitig, practical_QEM} which removes non-unitarity of gates by randomly inserting recovery
operations into the circuit.
Despite the close formal connection, PAI is quite different conceptually,
e.g., all gates involved in PAI are unitary, PAI does not apply gate insertions but rather
applies the same gate at different angle settings.
Furthermore, the quasiprobability decomposition
in PAI in \cref{eq:quasiprob} is known by construction as opposed to the experimentally learned
approximate models in PEC~\cite{strikis2021learning, PhysRevResearch.3.033098, berg2022probabilistic, montanaro2021error, cai2022quantum}.

\begin{figure}[tb]
	\begin{centering}
		\includegraphics[width=0.45\textwidth]{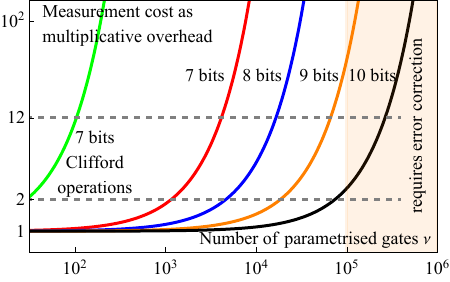}
	\end{centering}
	\caption{ 
		The measurement cost of PAI 
		is increased compared to the case when one has access to
		continuous rotation angles, see  \cref{eq:measurement_overhead}. 
		(solid lines) Worst-case measurement overhead $\lVert g  \rVert_1^2$ of PAI as a function of the number
		of parametrised gates in the quantum circuit. The number of gates one can reasonably
		(with an overhead at most $12$) implement with PAI is approximately $2^{2 (B_{min} -1)}$
		where $B_{min}$ is the number of bits used to digitise the rotation angle in \cref{fig:fig1}.
		As these estimates rely on worst-case bounds, we observe in numerical simulations
		that the actual number of gates can be significantly larger.
		(red vs. green lines) our optimal scheme can achieve many orders of magnitude smaller overheads
			than prior techniques based on Clifford operations~\cite{suguru_fault_tolerant}.
			(yellow region) very deep circuits will require quantum error correction and thus even for
			late-NISQ devices no more than 9 bits will be necessary.
		\label{fig:overhead}
	}
\end{figure}

\subsection{Estimating expected values~\label{sec:expval}}
Typical near-term and early fault-tolerant quantum algorithms use quantum computers for estimating 
expected values $o = \tr[O \ucirc |0\rangle\langle0|]$
of an observable $O$~\cite{cerezoVariationalQuantumAlgorithms2021a, endoHybridQuantumClassicalAlgorithms2021, bharti2021noisy}.
Thus one applies a state-preparation circuit to a fixed
reference state as $\ucirc |0\rangle\langle0|$;
One then performs a measurement whose outcome is generally a
random variable; By averaging over many repeated measurement outcomes one
obtains an empirical estimate of the expected value $o$. 
Without loss of generality we assume a normalised observable
$\lVert O \rVert_\infty$ and thus the number of repetitions required to determine the expected value to
a precision $\epsilon$ scales as $\nshot \leq \epsilon^{-2}$.

\begin{figure*}[t]
	\begin{centering}
		\includegraphics[width=\textwidth]{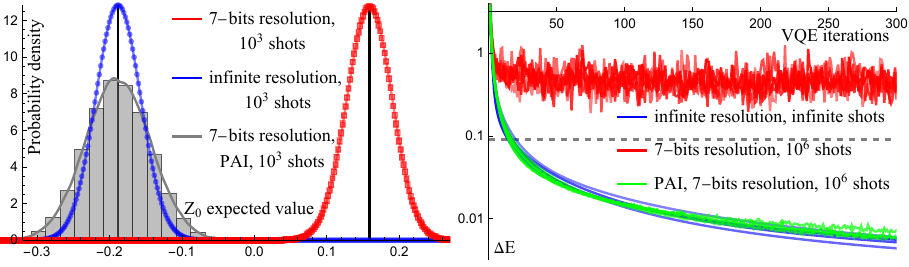}
	\end{centering}
	\caption{ 
		(left)
		Distribution of estimated expected values $\langle Z_0 \rangle$
		using $1000$ circuit repetitions (shots)
		in a deep, $12$-qubit trotter circuit of $l=50$ layers that consists of $\nu = 1786$ parametrised gates.
		(left, red) using the nearest notch settings at $7$ bits resolution
		results in a shifted mean (black vertical line) due to over/under rotations.
		Experimentally estimated histogram (grey) of PAI at $7$ bits resolution
		is centred around the same mean as the ideal one (blue) assuming infinite resolution
		but its distribution width is slightly increased.
		(right)
		Energy distance $\Delta E$ from the ground-state during a
		gradient descent search of a $12$-qubit spin-ring problem
		(energy shown assuming infinite resolution to inform about the optimiser's progress).
		A relatively deep circuit of $\nu =540$ parametrised gates is used.
		Gradient estimation is performed with (blue) infinite rotation-angle resolution
		and infinite number of shots;
		(red) using only the nearest notch settings at $7$ bits of resolution and $10^6$ shots;
		(green) using PAI with $10^4$ shots at only $100$ different circuit configurations.
		PAI (green) significantly outperforms the naive approach (red) despite
		it uses the same amount of quantum resources and
		essentially recovers the performance of the ideal optimiser (blue).
		Additionally shown is the energy (dashed grey)
		at the notch settings nearest to the ground-state parameters.
		\label{fig:numerics}
	}
\end{figure*}

Since we assume access to only discrete rotations, whenever the
hardware is instructed to execute the state-preparation circuit $\ucirc$ 
the parametrised gates are replaced by one of the discrete
circuit variants $\mathcal{U}_{\underline{l}}$. After performing a measurement,
one multiplies the random outcome with a factor $ \lVert g  \rVert_1  \sign(g_{\underline{l}} )$
that can have negative signs. 
As a consequence, the variance of the estimator is magnified which implies an increased number
of circuit repetitions.
\begin{statement}\label{stat:variance}
	Applying PAI to the estimation of an expected value
	results in an unbiased estimator $\hat{o}$ of the expected value of an observable
	as $\mathbb{E}[\hat{o}] =  \tr[O \ucirc |0\rangle\langle0|] = o$.
	The number of repetitions required to determine the expected value $o$ to accuracy $\epsilon$
	scales as
	\begin{equation}\label{eq:measurement_overhead}
			\nshot \leq  \epsilon^{-2}	\lVert g  \rVert_1^2 
		= \epsilon^{-2} \prod_{j=1}^\nu \lVert \gamma^{(j)}(\theta_j)  \rVert_1^2,
	\end{equation}
where $\lVert g  \rVert_1$ is simply a product of the single-gate norms $\lVert \gamma^{(j)}(\theta_j)  \rVert_1$
from \cref{stat:one_gate}.
\end{statement}
Indeed, to achieve the same precision, PAI has an increased measurement cost
compared to having physical access to continuously parametrised gates.
In the worst case, when all gate angles are exactly halfway between two notches
as $\theta_j = \Delta_j/2$,
this overhead scales as $e^{\nu \Delta_{max}^2/4}$ where $\Delta_{max}$ is the largest discretisation
across the different parametrised gates. 

The overhead is actually quite reasonable as long as the exponent does not
significantly exceed $1$ as illustrated in \cref{fig:overhead}. 
Thus, in order for the circuit repetitions to not exceed a
$12$-fold increase (grey dashed line in \cref{fig:overhead}),
the number $\nu$ of parametrised gates in a circuit that can be implemented with PAI
is limited by the lowest resolution $B_{min}$ of the gate discretions as
$\nu \leq  2^{2 (B_{min} -1)}$ .

For example, at $B=7$ bits resolution $4096$ gates can still reasonably be implemented,
while $10$ bits resolution allows over a quarter of a million gates, which is certainly sufficient for most near-term applications~\cite{cerezoVariationalQuantumAlgorithms2021a, endoHybridQuantumClassicalAlgorithms2021, bharti2021noisy}.
Furthermore, one can significantly reduce these costs by ``turning off''
PAI for the gates that are not contained in the light cone of the analysed observable $O$~\cite{pec_shadow_2023,locality_error_mitigation_ibm_2023}.

\section{Numerical simulations}
We consider a typical practical benchmarking task of simulating the spin-ring Hamiltonian
\begin{equation}\label{eq:spin_ring_hamiltonian}
	\mathcal{H} = \sum_{k \in \text{ring}(N)} \omega_k Z_k + J \Vec{\sigma}_k\cdot\Vec{\sigma}_{k+1},
\end{equation}
with coupling $J = 0.3$ and uniformly random $-1 \leq \omega_k \leq 1 $.
This spin problem is relevant in condensend-matter physics in understanding many-body localisation
in which early quantum computers might be very useful~\cite{nandkishore_2015,childs_2018, luitz_2015}.

\emph{Time evolution}---We first consider simulating time evolution
which is one of the most natural applications of quantum
computers~\cite{mcardle2020quantum, cao2019quantum, shadow_spec_2023}
and focus on Trotterisation that is a commonly applied simulation technique~\cite{berry2007efficient}; 
it approximates the time evolution $e^{-i t \mathcal{H} }$
as repeated layers of evolutions under the individual Hamiltonian terms for small times $\delta t$. Since the evolution under
each Hamiltonian term in \cref{eq:spin_ring_hamiltonian} is a Pauli rotation gate, e.g., $\mathcal{R}(\omega_k \delta t )$,  a layer of the time evolution
circuit is just a series of rotation gates each tuned to its relevant small rotation angle;
this layer is then repeated a large number $t/\delta t$ of times.

Rounding the rotation angles to nearest notch settings, e.g., $\omega_k \delta t  \mapsto \Theta_1$,
leads to a significant coherent error
as it implements incorrect evolution times and/or incorrect interaction terms $\omega_k$, thus
near phase transitions the discrepancy might be radical.
Measuring an expected value with a fixed number of shots leads to a biased probability distribution
as we illustrate in \cref{fig:numerics}(left, red).
In contrast, PAI results in a probability distribution that is centred around the exact mean
in \cref{fig:numerics}(left, blue and grey) while its distribution width is slightly increased.
The increase in width is actually significantly lower than
our worst-case estimates in \cref{stat:variance}
and we quantify in the Appendix that indeed generally this is the case when all rotation angles are small or are
close to one of the notch settings as in case of trotterised time-evolution circuits.

\emph{Finding eigenstates}---We next consider finding eigenstates of the Hamiltonian in \cref{eq:spin_ring_hamiltonian};
A broad range of techniques are available in the literature including ones that target near-term
and early fault-tolerant quantum computers~\cite{cerezoVariationalQuantumAlgorithms2021a, endoHybridQuantumClassicalAlgorithms2021, bharti2021noisy, covar,van2021measurement, koczor2020quantumAnalytic, PhysRevA.106.062416}.
We use the same trotterised circuit structure as we used for time evolution but
we optimise the angles of the rotation gates so that the energy $\tr[\mathcal{H} \ucirc |0 \rangle\langle 0|]$
of the emerging state is minimal -- this variant of the variational quantum eignesolver
uses the Hamiltonian Variational Ansatz as in case of QAOA~\cite{farhi2014quantum,cerezoVariationalQuantumAlgorithms2021a, endoHybridQuantumClassicalAlgorithms2021, bharti2021noisy}.

\cref{fig:numerics} (right, red lines) illustrates that a gradient descent optimiser does not manage to
meaningfully lower the energy when the gradient is calculated using only nearest notch settings
due to the coherent discrepancy in the output state.
In contrast, using the same quantum resources (same number of circuit repetitions and discretised gates)
but estimating the gradient using PAI
matches the performance of an ideal quantum circuit that has infinite angular resolution in
\cref{fig:numerics} (right, green and blue lines). 
We also note that formally our PAI protocol applies a different, randomly chosen circuit variant at each circuit repetition.
However, reconfiguring circuits will likely be a bottleneck for some quantum hardware platforms
and thus it is desired to run the same circuit variant multiple times.
Indeed, \cref{fig:numerics} (right, green) only uses $100$ different circuit variants---each of which is 
repeated $10^4$ times---which demonstrably does not compromise the optimiser's performance.

\section{Generalisations\label{sec:gen}}
A number of generalisations and further applications of our approach are apparent.

First, our results in \cref{theo1} directly apply to settings where the discretisation in
\cref{fig:fig1} is not uniform:
Such a non-uniform discretisation of angular settings may arise from a non-linear relationship between the
control field amplitude and the rotation angle achieved, for example when modulating the exchange
interaction~\cite{SymExchange} between two spin qubits or applying a Stark shift~\cite{StarkGary}.

Second, the approach can be generalised to quantum gates beyond gate generators
of eigenvalues $\pm 1$.
One then writes a system of equations similar to \cref{eq:quasiprob} but uses
more variables and more discrete gate angle settings, and solves for the variables
either analytically or numerically.

Third, as we detail in \cref{app:approximate}, a variant of PAI would be to simply omit the rarely-occurring third, antipolar angle from the choices and instead select between only the nearest notches. Conceivably this would offer a slight simplification of hardware, however this would fundamentally limit the device's capacity to obtain expectation values that are unbiased with respect to the values obtained from an ideal (continuous angle) system.

Fourth, in our spin-ring simulations we assumed standard trotterisation is used, however,
for quantum chemistry Hamiltonians one may significantly benefit from
randomised compilation techniques, such as qDRIFT~\cite{campbell2019random}.
As we now detail, PAI can be seamlessly combined with such randomised compilers.
In particular, given the Hamiltonian $\mathcal{H} = \sum_{q=1}^L h_q \mathcal{H}_q$ with
coefficient $\ell^1$ norm $\lambda = \sum_{q=1}^L |h_q|$, a standard
trotter circuit of $r$ steps consists overall of $N=Lr$ gates. 
In qDRIFT one randomly chooses a sequence of the rotation gates  $R_q(\tau) = e^{i \tau \mathcal{H}_q/2}$ with the benefit
that the sequence length depends on the norm $\lambda$ rather than on the number of terms $L$
and that each gate in the sequence needs only have a fixed, constant
rotation angle $\tau = 2t \lambda /N$.
These angles are indeed potentially small and the application of PAI is just
as relevant as in the case of standard trotterisation: the combination with PAI
proceeds by first randomly choosing a gate $\mathcal{R}_q(\tau)$ according to qDRIFT and
then implementing the continuous rotation angle $\theta=\tau$ by
randomly choosing one of the relevant notch settings $\mathcal{R}_q(\Theta_k)$ 
according to PAI.

Finally, we note that the present approach is not limited to near-term applications
and might also be useful in the early fault-tolerance regime.
In particular, ref~\cite{suguru_fault_tolerant} considered the Solovay-Kitaev decomposition whereby
one aims to approximate a continuous rotation $\mathcal{R}(\theta)$ as
a sequence of Clifford and $T$ gates. Ref~\cite{suguru_fault_tolerant} then considered
PEC to mitigate the approximation error of this decomposition by randomly applying Clifford recovery operations
and already noted it leads to a significant measurement overhead.
Indeed, our general solution in \cref{theo2} can be applied to 
Clifford recovery operations
$\mathcal{R}(\pi/2)$ and $\mathcal{R}(\pi)$ by substituting the (suboptimal) angles $A= \pi/2$, $B= \pi$ and $\theta=\Delta/2$.
Given in our approach we choose $A$ and $B$ optimally, our scheme achieves many orders of magnitude lower
measurement overhead in typical practical scenarios as illustrated in \cref{fig:overhead}(green vs. red lines).

\section{Discussion and Conclusion\label{sec:discussion}}

We present PAI which effectively upgrades the capabilities of a quantum hardware that can only realise discrete
rotation angles to a device that can perform arbitrary, continuous rotation angles.
We achieve this by randomly choosing one of three possible rotation angles  in each parametrised gate
such that on average the exact, desired unitary rotations are performed.
The limitation of only being able to perform a discrete set of rotation angles manifests itself
in an increased number of circuit repetitions when measuring expected values of observables.
	
We upper bound this measurement overhead and conclude it is negligible
even for circuits consisting of hundreds or a few thousand parametrised
gates at a resolution as low as $B=7$ bits.
Apart from the slightly increased number of circuit repetitions, the
present approach requires no additional quantum resources.
PAI can be compared with a number of well-established prior techniques.

\emph{Cooperative optimum control}---it was proposed in \cite{braun2010cooperative} that
shaped pulses $\mathcal{R}_k$ that implement a desired rotation
need not be individually accurate but rather need only satisfy a relaxed
condition that the average of a series of pulse variants
are required to be accurate over many repeated measurement rounds.
The present approach is indeed quite comparable, however, we allow for the additional freedom
that different gate/circuit variants are weighted. For example, in a single notched gate implementation
we have three gate variants in \cref{eq:quasiprob} but the third one $\mathcal{R}(\Theta_{k}{+}\pi)$
is only rarely applied due to its small probability $\mathcal{O}(\Delta^2)$.
We expect, however, that the present techniques will enable new developments
in exploiting pulse-optimisation techniques for near-term applications \cite{meitei2021gate,de2023pulse}.

\emph{Quantum Error Mitigation}---techniques~\cite{cai2022quantum} can be combined with the present approach seamlessly
	in order to mitigate experimental imperfections of the gates $\mathcal{R}_1$, $\mathcal{R}_2$
	and $\mathcal{R}_3$ in \cref{eq:rots} as we detail in \cref{app:qem}.
	The measurement overhead of PAI from \cref{sec:expval} then increases
	to $e^{\nu (4 \epsilon + \Delta_{max}^2/4)}$,
	thus the resolution $\Delta_{max}$ should be engineered in accordance with the
	error rates $\epsilon$ of the physical gates via $4 \epsilon \approx \Delta_{max}^2/4$.
	Specifically, for first generations of devices ($10^{-3} \leq \epsilon \leq 10^{-2}$)
	$B=5$ digits of precision may suffice while one needs $B=6-7$ bits of precision
	for NISQ devices in the early practical quantum advantage regime ($10^{-4} \leq \epsilon \leq 10^{-3}$).
	
	Furthermore, our approach leverages on a similar quasiprobability decomposition to PEC~\cite{gambetta_error_mitig,practical_QEM, cai2022quantum},
	and we can thus take advantage of
	a rich literature to, e.g., use light-cone arguments  to significantly reduce sampling costs~\cite{pec_shadow_2023,locality_error_mitigation_ibm_2023,foldager2023can},
	while PAI can also be immediately combined with classical shadows
	via~\cite{pec_shadow_2023,shadow_huang_2020}.

\emph{Circuit Knitting}---Related quasiprobability decompositions have been used for
replacing two-qubit entangling gates with classically post processed random
implementations of single-qubit operations~\cite{mitarai2021constructing,piveteau2022circuit} which was
termed Circuit knitting.
The approach has a cost $\mathcal{O}(9^\nu/\epsilon^2)$ for implementing $\nu$ two-qubit gates
and is thus limited to only implementing a few, e.g., 3-4, quantum gates whereas the present
protocol works well in the regime of tens of thousands of gates assuming resolutions in the range $7-9$ bits.

To conclude, we expect the present technique will be an important and useful tool
in designing optimal quantum hardware: Our analysis suggests that first generation
quantum hardware, being practically limited to only a few thousand gate operations
(e.g., due to limited coherence times), will need no more than $7$ bits of resolution in the control systems.
As the technology matures, future generations of hardware are expected to be able
to execute tens of thousands of quantum gates without error correction which still, however,
requires no more than $9$ bits of angular resolution.

\section*{Acknowledgments}
B.K. thanks the University of Oxford for
a Glasstone Research Fellowship and Lady Margaret Hall, Oxford for a Research Fellowship.
The numerical modelling involved in this study made
use of the Quantum Exact Simulation Toolkit (QuEST), and the recent development
QuESTlink~\cite{QuESTlink} which permits the user to use Mathematica as the
integrated front end, and pyQuEST~\cite{pyquest} which allows access to QuEST from Python.
We are grateful to those who have contributed
to all of these valuable tools. 
The authors would like to acknowledge the use of the University of Oxford Advanced Research Computing (ARC)
facility~\cite{oxford_arc} in carrying out this work
and specifically the facilities made available from the EPSRC QCS Hub grant (agreement No. EP/T001062/1).
The authors also acknowledge funding from the
EPSRC projects Robust and Reliable Quantum Computing (RoaRQ, EP/W032635/1)
and Software Enabling Early Quantum Advantage (SEEQA, EP/Y004655/1).

\appendix

\section{Single rotation gate\label{app:coeffs}}
As they form one-parameter groups, any overrotated quantum gate can be written as
\begin{equation}\label{eq:overrot}
	\mathcal{R}(\Theta_k + \theta) = \mathcal{R}(\Theta_k) \mathcal{R}(\theta),
\end{equation}
where $0 \leq \theta \leq \Delta$ is a small, arbitrary overrotation.
Due to this property, we need only expand the rotation $\mathcal{R}(\theta)$
into a linear combination of rotations at different angles.

Focusing on Pauli gates of the form $e^{-i \theta P_k/2}$ for
any Pauli string $P_k \in \{ \mathrm{Id}, X, Y, Z \}^{\otimes N}$
in an $N$-qubit system,
ref.~\cite{koczor2022quantum} showed that any Pauli rotation can be 
decomposed as a linear combination of the same gate at different rotation angles.
We also note that any gate generator with eigenvalues $\pm1$ is covered by our
formalism, e.g., our results similarly apply to parametrised SWAP gates.
The decomposition follows as
\begin{align}\nonumber
	\mathcal{R}(\theta) &= 
	\frac{ 1+\cos\theta }{2} \mathcal{R}(0)
	+
	\frac{\sin\theta }{2} [ \mathcal{R}(\pi/2) - \mathcal{R}(-\pi/2)]\\
	&+
	\frac{ 1-\cos\theta }{2} \mathcal{R}(\pi). \label{eq:channel_form}
\end{align}
We use the above decomposition to obtain the quasiprobabilities $\gamma_l$ in \cref{eq:quasiprob}.
In particular, by combining \cref{eq:channel_form} with \cref{eq:quasiprob} we obtain the following non-linear system of equations
\begin{align}
	\nonumber
&\gamma_1 [1{+}\cos(0)] {+} \gamma_2 [1{+}\cos(\Delta)] {+} \gamma_3 [1{+}\cos(\pi)] = 1{+}\cos(\theta),\\ 
&\gamma_1 \sin(0) {+} \gamma_2 \sin(\Delta) {+} \gamma_3 \sin(\pi) = \sin(\theta),\label{eq:eqsys}\\
&\gamma_1 [1{-}\cos(0)] {+} \gamma_2 [1{-}\cos(\Delta)] {+} \gamma_3 [1{-}\cos(\pi)] = 1{-}\cos(\theta).
\nonumber
\end{align}
The above system of equations is solved by the set of coefficients
$\gamma_1$, $\gamma_2$ and $\gamma_3$ as
\begin{align}\label{eq:solution}
	\gamma_1 =&
	\csc \left(\frac{\Delta}{2}\right) \cos \left(\frac{\theta }{2}\right) \sin \left(\frac{\Delta}{2}-\frac{\theta }{2}\right),\\ \nonumber
	\gamma_2 =&
	\csc (\Delta) \sin (\theta ),\\ \nonumber
	\gamma_3 =&
	-\sec \left(\frac{\Delta}{2}\right) \sin \left(\frac{\theta }{2}\right) \sin \left(\frac{\Delta}{2}-\frac{\theta }{2}\right).
\end{align}
We can also analytically compute the vector norm as
\begin{equation}\label{eq:l1_norm}
\lVert \gamma\rVert_1 =	\sec \left(\frac{\Delta}{2}\right) \cos \left(\frac{\Delta}{2}-\theta \right).
\end{equation}

Finally, we verify that the estimator of the channel in \cref{stat:one_gate} is unbiased
by substituting the probability
$p_l(\theta) = |\gamma_l(\theta)| / \lVert \gamma(\theta) \rVert_1$ as
\begin{align*}
	\mathbb{E}[ \hat{\mathcal{R}}(\Theta_k + \theta)]
	=&  \sum_{l = 1}^3  	\lVert \gamma(\theta) \rVert_1  p_l(\theta) \mathrm{sign}[ \gamma_l(\theta) ]   \mathcal{R}_l\\
	= & \sum_{l = 1}^3    \mathrm{sign}[ \gamma_l(\theta) ]  |\gamma_l(\theta) |  \mathcal{R}_l\\
	=&  \sum_{l = 1}^3    \gamma_l(\theta)  \mathcal{R}_l
	= \mathcal{R}(\Theta_k + \theta)
\end{align*}
where the last equation follows from \cref{eq:quasiprob}.

\section{Asymptotic expansions}

We compute asymptotic expressions for small $\Delta$
by introducing the ratio $\lambda = \theta/\Delta$ as
\begin{align} \label{eq:expansion}
	\gamma_1 	&= (1-\lambda ) + \frac{   \lambda (1- \lambda)  (1 - 2 \lambda)   }{12} \Delta^2  + \mathcal{O}(\Delta^4),\\ \nonumber
	\gamma_2	&=
	\lambda + \frac{ (\lambda - \lambda^3)  }{6} \Delta^2 + \mathcal{O}(\Delta^4), \\ \nonumber
	\gamma_3 &= - \frac{ \lambda (1 - \lambda)  } {4}   \Delta^2   + \mathcal{O}(\Delta^4).
\end{align}
Thus the probabilities can be expanded as
\begin{align*}
	p_1 &= 1-\lambda  + \mathcal{O}(\Delta^2),\\
	p_2 &= \lambda + \mathcal{O}(\Delta^2),\\
	p_3 &=  \frac{\lambda(1-\lambda)}{4} \Delta^2 + \mathcal{O}(\Delta^4).
\end{align*}
Similarly, later we will make use of the squared norm $\lVert \gamma(\theta) \rVert^2_1$ 
and its expansion into leading terms as
\begin{equation}\label{eq:norm}
	\lVert \gamma(\theta) \rVert_1^2 =  
	1 + \lambda (1- \lambda) \Delta^2 
	+ \mathcal{O}(\Delta^4).
\end{equation}
We can also generally upper bound  this norm by its worst case value at $\lambda = 1/2$ 
and substitute the above expansion as
\begin{equation}\label{eq:normbound}
	\lVert \gamma(\theta) \rVert^2_1  \leq \lVert \gamma(\Delta/2) \rVert^2_1   = 1 + \Delta^2/4 + \mathcal{O}(\Delta^4).
\end{equation}

\begin{figure*}[t]
	\begin{centering}
		\includegraphics[width=0.45\textwidth]{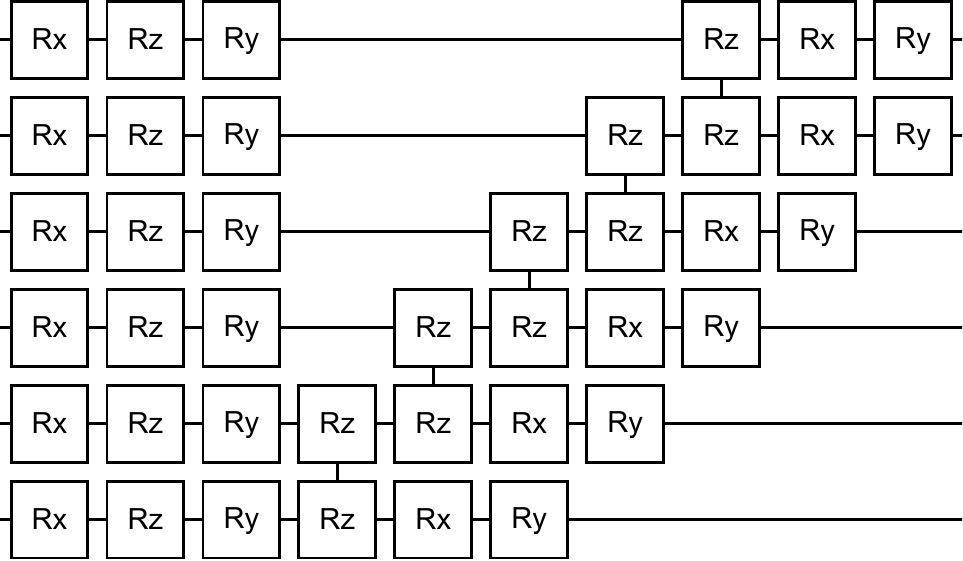}
		\includegraphics[width=0.45\textwidth]{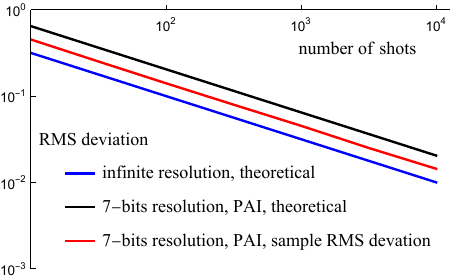}
	\end{centering}
	\caption{ 
		(left) A single layer of the ansatz structure used in our simulation show for $6$ qubits.
		(right) Same experiment as in \cref{fig:numerics} showing how the  root mean square (RMS) deviation decreases as we increase the number of shots.
		(blue) standard shot noise limit using the device of infinite resolution; (black) sample RMS deviation -- the ideal device's standard
		deviation is slightly increased by  PAI as black is slightly above blue (red) our theoretical worst-case bound based on the variance
		of PAI is well above the sample RMS deviation.
		\label{fig:errorPlot}
	}
\end{figure*}

\section{Circuit construction \label{app:unbiased}}
We now present a detailed derivation of the estimator of the full circuit in \cref{stat:circ}
and prove its unbiasedness.
To simplify notations we assume the circuit contains no other gates than the parametrised ones as
\begin{equation}\label{eq:circ}
	\ucirc = \prod_{j=1}^\nu  \mathcal{R}^{(j)}(\Theta_{k_j} + \theta_j),
\end{equation}
however, it is straightforward to generalise to the case when the circuit contains other
non-parametrised gates too.
Here $\mathcal{R}^{(j)}$ denotes the $j^{th}$ parametrised gate which is ideally set to the 
continuous angle that we express as an over rotation by an
angle $\theta_j$ relative to the notch setting $\Theta_{k_j}$.
Let us, similarly to \cref{eq:rots}, denote the discrete rotations as
\begin{align*}
	\mathcal{R}^{(j)}_1 :=& \mathcal{R}^{(j)}(\Theta_{k_j}), \quad \mathcal{R}^{(j)}_2 := \mathcal{R}^{(j)}(\Theta_{k_j+1}), \\
	\mathcal{R}^{(j)}_3 :=& \mathcal{R}^{(j)}(\Theta_{k_j} {+} \pi).
\end{align*}
Using \cref{stat:one_gate} we replace each gate in \cref{eq:circ} with its estimator as
\begin{align*}
	\ucirce 
	=& \prod_{j=1}^\nu  \hat{ \mathcal{R} }^{(j)}(\Theta_{k_j} + \theta_j)\\
	=& \prod_{j=1}^\nu  \lVert \gamma^{(j)}(\theta_j) \rVert_1   \mathrm{sign}[ \gamma^{(j)}_{l_j}(\theta_j) ] \mathcal{R}_{l_j}^{(j)}\\
	=& \lVert g  \rVert_1  \sign(g_{\underline{l}} ) \prod_{j=1}^\nu   \mathcal{R}_{l_j}^{(j)}\\
	=& \lVert g  \rVert_1  \sign(g_{\underline{l}} ) \mathcal{U}_{\underline{l}}.
\end{align*}
Above we have introduced the abbreviation $\mathcal{U}_{\underline{l}} := \prod_{j=1}^\nu  \mathcal{R}_{\underline{l}}^{(j)} $
for the discrete-angle circuit variants where the multi index $\underline{l}:= (l_1, l_2, \dots  l_\nu) \in 3^\nu$
collects all discrete angle settings.
We have also defined the coefficient
\begin{equation}\label{eq:coeffs}
	g_{\underline{l}} = \prod_{j=1}^\nu  \gamma^{(j)}_{l_j}(\theta_j).
\end{equation}
The norm of this coefficient vector is simply a product as $\lVert g  \rVert_1 = \prod_{j=1}^\nu \lVert \gamma^{(j)}  \rVert_1$
and thus the associated probability of choosing the circuit variant is $p_{\underline{l}} = |g_{\underline{l}}| /\lVert g  \rVert_1 $.

One can verify that indeed the estimator is unbiased in the sense that
\begin{align*}
		\mathbb{E}[ \ucirce ]
		=&		\sum_{\underline{l}} p_{\underline{l}} \times \Big[ \lVert g  \rVert_1  \sign(g_{\underline{l}} ) \mathcal{U}_{\underline{l}} \Big]\\
		=& 	\sum_{\underline{l}}  |g_{\underline{l}}|  \sign(g_{\underline{l}} ) \mathcal{U}_{\underline{l}}
		= 	\sum_{\underline{l}}  g_{\underline{l}}  \mathcal{U}_{\underline{l}}		
\end{align*}
By substituting the definition of $ g_{\underline{l}} $ and $\mathcal{U}_{\underline{l}}$ as
\begin{align*}
			\mathbb{E}[ \ucirce ]
	=& \sum_{\underline{l}} \Big( \prod_{j=1}^\nu  \gamma_{l_j}(\theta_j) \Big) \Big( \prod_{j=1}^\nu  \mathcal{R}_{\underline{l}}^{(j)} \Big)\\
	=& \prod_{j=1}^\nu \Big(    \sum_{l_j =1}^3    \gamma_{l_j}(\theta_j) \mathcal{R}_{\underline{l}}^{(j)}   \Big)\\
	=& \prod_{j=1}^\nu 	 \mathcal{R}^{(j)}(\Theta_{k_j} + \theta_j)	= \ucirc,
\end{align*}
one indeed finds the correct expected value, where
in the second to last equation we used \cref{eq:quasiprob}.

\subsection{Expected values}
We consider the estimation of an arbitrary observable. We model the
measurement process using a Positive Operator Valued Measure $E = \{E_b\}$
which is a collection of positive operators $E_b$, called the effects. 
The simplest case is the case of perfect projective measurements
via the effects $E_b = | b \rangle\langle b|$ where $b \in \{0,1\}^N$
are bitstrings. The effects can be more general positive operators
which is useful when modelling measurement errors but in general $\sum_b E_b = \openone$.

The expected value of an arbitrary
observable $O$ is then estimated via the estimator $\hat{x} = \tr[O E_b]$
and the expectation value is $\mathbb{E}[\hat{x}] = \sum_b  q_b \tr[O E_b]$.
Generally, the probability of observing one of the outcomes in 
an arbitrary state $\rho$ is $q_b = \tr[ \rho E_b]$.
Indeed for the case of projective measurements in the eigenbasis of $O$
one obtains the usual
\begin{equation*}
	\mathbb{E}[\hat{x}] =
	\sum_b  \tr[O q_b E_b] = \tr\Big[ \rho \sum_b  \tr[O E_b]  E_b\Big] 
 = \tr[ \rho O],
\end{equation*}
where we used the equality $O = \sum_b  \tr[O E_b]  E_b$.

Let us now assume that the state is prepared via the circuit as $ \ucirc |0\rangle\langle 0|$
and our aim is to estimate the expected value $\tr[  O \ucirc |0\rangle\langle 0|]$.
As described in \cref{stat:circ}, we choose randomly and run the circuit variants $\mathcal{U}_{\underline{l}}$
and we multiply the individual outcomes with the relevant prefactor $\lVert g  \rVert_1 \sign(g_{\underline{l}} )$
thereby estimating $\lVert g  \rVert_1 \sign(g_{\underline{l}} ) \tr[  O \mathcal{U}_{\underline{l}} |0\rangle\langle 0|]$.
Formally, this results in the estimator as 
\begin{equation}\label{eq:estimator_obs}
	\hat{o} =   \lVert g  \rVert_1 \sign(g_{\underline{l}} )   \tr[O E_b],
\end{equation}
in which we multiply the individual outcomes $\tr[O E_b]$ with the relevant prefactors.
The probability of observing an outcome is $q_{\underline{l}, b} =  p_{\underline{l}} \tr[ E_b\,  \mathcal{U}_{\underline{l}} |0\rangle\langle 0| ]$, where $p_{\underline{l}}$ is the probability of choosing the circuit variant $\mathcal{U}_{\underline{l}}$.
One can indeed verify that we obtain the correct expected value as
\begin{align*}
	\mathbb{E}[ \hat{o} ] =& \sum_{b,\underline{l}}  q_{\underline{l}, b} \times  \lVert g  \rVert_1  \sign(g_{\underline{l}} ) \tr[O E_b]\\
	=&  \sum_{b,\underline{l}}  p_{\underline{l}}   \, \tr[ E_b \, \mathcal{U}_{\underline{l}} |0\rangle\langle 0| ] \, \times \,
	 \lVert g  \rVert_1  \,  \sign(g_{\underline{l}} ) \, \tr[O E_b]   \\
	 =& \sum_{b} \sum_{\underline{l}}  g_{\underline{l}}   \, \tr[ E_b \,  \mathcal{U}_{\underline{l}} |0\rangle\langle 0| ]      \tr[O E_b] \\
	  =& \sum_{b} \tr[ E_b \, \Big( \sum_{\underline{l}}  g_{\underline{l}}  \mathcal{U}_{\underline{l}} \Big) |0\rangle\langle 0| ]      \tr[O E_b]\\
	  =&  \tr[ \Big(\sum_{b} \tr[O E_b] E_b \Big) \, \ucirc |0\rangle\langle 0| ]      
\end{align*}
where we used that $  \lVert g  \rVert_1   p_{\underline{l}} = |g_{\underline{l}} |$
and substituted that $ \sum_{\underline{l}}  g_{\underline{l}}  \mathcal{U}_{\underline{l}}  =  \ucirc$.
Indeed we obtain the expected value $	\mathbb{E}[ \hat{o} ] = \tr[  O \ucirc |0\rangle\langle 0|]$.

\subsection{Variance of estimators}
We consider the variance of the estimator of the quantum-mechanical expected value in \cref{eq:estimator_obs} as
\begin{equation*}
	\var[\hat{o}] = \var \Big[  \lVert g  \rVert_1  \sign(g_{\underline{l}} ) \tr[O E_b] \Big],
\end{equation*}
for any observable $O$.
Recall that we calculate the variance as $\var [\hat{o}] = \mathbb{E}[(\hat{o} - \mathbb{E}[\hat{o}])^{2}]$
and since our estimator $\mathbb{E}[\hat{o}] = \tr [  O	\ucirc \rho ]$ is unbiased
we need only upper bound the term $\mathbb{E}[\hat{o}^{2}] $ as
\begin{align*}
	\var  [\hat{o}] 
	&\leq  	\mathbb{E}[\hat{o}^{2}]\\
	&=  \sum_{b,\underline{l}}  q_{\underline{l}, b} \times  \Big[ \lVert g  \rVert_1  \sign(g_{\underline{l}} ) \tr[O E_b] \Big]^2\\
	&=  \lVert g  \rVert_1^2 \sum_{b,\underline{l}}  q_{\underline{l}, b} \times \tr[O E_b]^2
	\leq \lVert g  \rVert_1^2 \lVert O \rVert_\infty^2.
\end{align*}
Above we have used that $\tr[O E_b] \leq \lVert O \rVert_\infty^2$
and used that the probability distribution satisfies $\sum_{b,\underline{l}}  q_{\underline{l}, b} = 1$.

After repeating the single-shot procedure $\nshot$ times and calculating the mean of the
individual outcomes $\mathrm{mean}(\hat{o}_1, \hat{o}_2, \dots \hat{o}_{\nshot})$,
the variance $\epsilon^2$ of the empirical mean scales inversely proportionally
with $\nshot$. Thus, the sample complexity to achieve accuracy $\epsilon$ is
\begin{equation*}
	\nshot \leq  \var  [\hat{o}]  /\epsilon^2 = 	\lVert g  \rVert_1^2 \lVert O \rVert_\infty^2 /\epsilon^2.
\end{equation*}

\subsection{Asymptotic expansion of the variance~\label{app:variance}}
Without loss of generality we assume that $	\lVert O \rVert_\infty^2 = 1$.
We can further expand the variance by recalling that the coefficients $\lVert g  \rVert_1^2 $ factorise into a
product form in \cref{eq:coeffs} as
\begin{equation*}
	\var  [\hat{o}] 
	\leq   \prod_{j=1}^\nu \lVert \gamma^{(j)}(\theta_j)  \rVert_1^2.
\end{equation*}
We generally upper bound the norms in \cref{eq:normbound} as
$\lVert \gamma^{(j)}(\theta_j)  \rVert_1^2 \leq \lVert \gamma^{(j)}(\Delta_j/2)  \rVert_1^2$
using the worst-case scenario, maximal value which is attained
at exactly halfway between two notch settings as $\theta_j = \Delta_j/2$.
We can thus generally upper bound the coefficients and expand the bound for small notch discretisations $\Delta_j \leq \Delta_{max}$ as
\begin{align*}
	\var  [\hat{o}] 
	\leq &  \prod_{j=1}^\nu \lVert \gamma^{(j)}(\Delta_j/2)  \rVert_1^2\\
	= & \prod_{j=1}^\nu  [1 + \Delta_j^2/4 + \mathcal{O}(\Delta_{max}^4)]\\
	\leq &
	[1 + \Delta_{max}^2/4 + \mathcal{O}(\Delta_{max}^4)]^\nu\\
	\approx &  \, e^{ \nu \Delta_{max}^2/4 } 
\end{align*}
where $\Delta_{max}$ is the largest of the notch discretisations $\Delta_j$.
By substituting the explicit from of $\Delta_j$ for the discretisations from \cref{eq:delta} we obtain in the exponent $\pi^2 \nu 2^{-2B_{min}}$.
Thus the minimal number of bits $B_{min}$ required for the variance to still be bounded as $\var  [\hat{o}]  \leq e^{\pi^2/4}
 \leq 12 	$
the number of gates $\nu$ needs to satisfy
\begin{equation*}
	\nu \leq  2^{2 (B_{min} -1)}.
\end{equation*}

\begin{figure}[tb]
	\begin{centering}
		\includegraphics[width=0.45\textwidth]{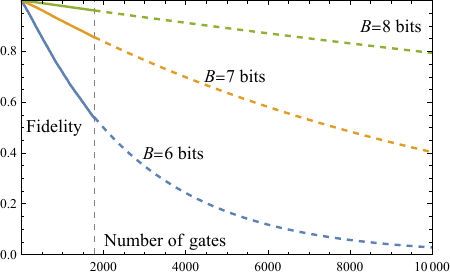}
	\end{centering}
	\caption{ 
		The same simulation of a time-evolution circuit as in \cref{fig:numerics}(left)
		but using our approximate scheme from \cref{app:approximate} whereby we only choose from
		the two nearest notch settings. This approach does not introduce a measurement overhead,
		however, the average gate implemented is non-unitary. The simulation shows that as we increase the number
		of gates to which we apply this approximate approach the fidelity of the output state drops exponentially.
		In comparison, at $B=7$ bits of precision and for $\nu = 1786$ parametrised gates we observe
		a $20\%$ increase in the distribution width in \cref{fig:numerics}(left) whereas here we observe a
		$15\%$ drop in the fidelity of the output state.
		We show extrapolated values for gatecounts larger than $\nu = 1786$.
		\label{fig:naive}
	}
\end{figure}

The above upper bound is attained only in the worst-case scenario when all rotation angles
are exactly halfway between two notch setting via $\lambda_j = 1/2$.
We can instead approximate the variance upper bound by expanding the vector norms using \cref{eq:norm} 
given the discretisation
are small in practice $\Delta_j \ll 1$ as
\begin{align*}
	\var  [\hat{o}] 
	= & \prod_{j=1}^\nu  [1 + \lambda_j (1- \lambda_j) \Delta_j^2 
	+ \mathcal{O}(\Delta_{max}^4)]\\
	= &\exp[ \sum_{j=1}^\nu \lambda_j (1- \lambda_j) \Delta_j^2 ] + \mathcal{O}(\nu \Delta_{max}^4)\\
	\leq &\exp[ \nu \tilde{\lambda} \Delta_{max}^2/4 ] + \mathcal{O}(\nu \Delta_{max}^4)
\end{align*}
Above we have introduced the notation 
for the average deviation of the rotation angles from the nearest notch settings as
\begin{equation*}
	\tilde{\lambda} = 4 \nu^{-1}\sum_{j=1}^\nu \lambda_j (1- \lambda_j),
\end{equation*}
which has the property $0 \leq \tilde{\lambda} \leq 1$.
We thus find that the minimum number of bits
in the discretisation needs to satisfy
\begin{equation*}
	\nu \leq \tilde{\lambda}^{-1} \, \,  2^{2 (B_{min} -1)}.
\end{equation*}
Indeed in the worst case $\tilde{\lambda} =1$ but when no rotation angle deviates from the nearest notch setting
by more than $25\%$ then $\tilde{\lambda} = 0.75$ and thus the maximal number of gates is increased inversely proportionally.

\section{Numerical simulations}

In \cref{fig:numerics}(left) we simulated a trotterised time-evolution circuit of our spin-ring problem at
12 qubits.
We used $l=50$ repeated small time evolution layers similar in structure to the ansatz circuit
shown in \cref{fig:errorPlot}(left) and the circuit overall consist of $\nu = 1786$ parametrised gates.  

We randomly generated $10^5$ instances of the circuit variants $\mathcal{U}_{\underline{l}}$
by randomly replacing the continuously parametrised gates with one of three discrete gate variants according
to the PAI protocol assuming $7$ bits resolution in all gates.
These circuits were simulated and $100$ shots were measured on qubit $0$ resulting in a list of $0$s and $1$s.
Overall, thus we had $10^7$ shots of the PAI protocol.

From these $10^7$ shots only $1000$ shots were uniformly randomly selected from which the expected value 
 $\langle Z_0 \rangle$ was calculated via the weighted, signed average of the PAI protocol, i.e., 
each shot yields a single-shot estimator of the expected value as $+1$ or $-1$ which we multiply with
the factor $ \lVert g  \rVert_1  \sign(g_{\underline{l}} )$ and calculate the mean of the $1000$ circuit repetitions.
This yields a single empirical estimate of  $\langle Z_0 \rangle$ from $1000$ shots.
We repeat this procedure $10^5$ times by always randomly choosing $1000$ shots from our pool
of $10^7$ shots. The empirical histogram in  \cref{fig:numerics}(left) was calculated from this
pool of empirical estimates of $\langle Z_0 \rangle$.
Additionally, \cref{fig:errorPlot}(right, red) shows the average root mean square deviation of the empirical 
estimate of $\langle Z_0 \rangle$ as an average of $10^4$ experiments each using $\nshot$ shots. Indeed
as we increase the number of shots the average deviation decreases as $\propto 1/\sqrt{\nshot}$.
The RMS deviation is below our theoretical worst-case upper bound \cref{fig:errorPlot}(right, black)
but it is increase compared to the case of infinite angular resolution \cref{fig:errorPlot}(right, blue).
In \cref{fig:numerics}(right) we simulated a deep Variational Hamiltonian Ansatz, a single layer of which is illustrated
in \cref{fig:errorPlot}(left). Our 12-qubit circuit consists overall of  $\nu =540$ parametrised gates.

\section{Approximate approach\label{app:approximate}}
We now define an approach whereby we omit the third notch setting $\Theta_k + \pi$
by effectively forcing the third expansion coefficient $\gamma_3 \rightarrow 0$ in \cref{eq:quasiprob}.
The approach has the advantage that it does not introduce a measurement overhead, however,
the approach results in an biased estimator of the parametrised quantum gate.

In particular, we choose the notch setting
$\Theta_k$ with probability $p_1 = \gamma_1 = 1-\lambda$ and we choose the notch
setting $\Theta_{k+1}$ with probability $p_2 = \gamma_2= 1-\gamma_1 = \lambda$ such that these non-negative
probabilities satisfy $\gamma_1 + \gamma_2 =1$.

The resulting average quantum gate is not only biased, but it is not unitary. As such, applying this
protocol to $\nu$ parametrised gates in a circuit yields an exponential decay of the target
fidelity as $e^{- \mathcal{O}(\Delta^2) \nu}$ since each gate has an error $\mathcal{O}(\Delta^2)$.
This is illustrated in \cref{fig:naive} where we apply the present protocol to the same 12-qubit
time-evolution circuit as in \cref{fig:numerics}(left). At $B=7$ bits of precision the fidelity drops by
$15\%$ when all $\nu = 1786$ gates are executed using the approximate approach.
In contrast, our PAI protocol yields perfect fidelity (mean value matches the ideal one in \cref{fig:numerics}(left))
but the probability distribution of measurement outcomes is increased by about a comparable factor, i.e., 
by about $20\%$. However, this increase in distribution width can be trivially overcome
just by repeating the measurement procedure by a slightly increased number of times.
In contrast a $15\%$ infidelity might imply a comparable error when measuring
expected values which would be prohibitive in practice.

However, if one still prefers to use the approximate approach then one can apply error mitigation techniques
to mitigate the drop in the fidelity. These indeed generally come at the cost of an increased sampling~\cite{cai2022quantum}
and may thus yield a worse performance than PAI. A straightforward approach would be to use the present approximate
scheme at $B$ and $B-1$ bits of precision, estimate expected values and extrapolate to the case of $\Delta \rightarrow 0$. 
While the individual expected values require no increased sampling, the shot noise in the extrapolated estimate
is increased and thus Zero Noise Extrapolation indeed requires overall an increased sampling.

\section{Optimality of the solution \label{app:optimal}}

It is clear from \cref{eq:channel_form} that the minimal number of rotation gates needed is three as
\cref{eq:channel_form} has three degrees of freedom, namely, $1\pm\cos(\theta)$ and $\sin(\theta)$.
While indeed we use 3 rotation gates, namely the two nearest notch settings and a polar opposite rotation,
we discuss in \cref{app:symm} that using more than 3 notch settings is suboptimal.

We now prove that our choice of 3 notch settings is optimal in the sense that any other
choice of 3 rotation angles yields a solution with a higher $\lVert \gamma \rVert_1$ norm --
which norm is of paramount importance for us to minimise.

\begin{theorem}\label{theo2}
	Given a target rotation gate as the quasiprobability decomposition (cf \cref{eq:quasiprob})
	\begin{equation}
		\mathcal{R}(\theta) = \gamma_1(\theta) \mathcal{R}_1 + \gamma_2(\theta) \mathcal{R}_2
		+  \gamma_3(\theta) \mathcal{R}_3,
	\end{equation}
	where the three rotation gates are chosen from any set of (possibly non-uniform) discrete notch settings that a machine can realise
	$\mathcal{R}_1, \mathcal{R}_2, \mathcal{R}_3 \in \{ \mathcal{R}(\Theta_q) \}$.
	Let us denote as $\Theta_k$ and $\Theta_{k+1}$ as the two nearest notch settings to $\theta$
	and define the distance $\Delta:= \Theta_{k+1} - \Theta_{k}$.
	The optimal choice of rotation gates
	that minimises $\lVert \gamma \rVert_1$ is the following
	\begin{equation*}
		\mathcal{R}_1 = \mathcal{R}(\Theta_k), \quad
		\mathcal{R}_2 = \mathcal{R}(\Theta_{k+1}), \quad
		\mathcal{R}_3 = \mathcal{R}(B),
	\end{equation*}
	where $B$ is the rotation angle nearest to $\Theta_{k} + \pi + \frac{\Delta}{2} $.
	In the present work we consider a uniform discretisation of $\Theta_k$, 
	thus there are two equivalent choices as $B=\Theta_{k} + \pi$ or as $B=\Theta_{k+1} + \pi$,
	as we detail in \cref{app:symm}.
\end{theorem}
\begin{proof}
As explained in \cref{eq:overrot}, without loss of generality the notch setting
can be defined as $\Theta_k := 0$. We then have two degrees of freedom that we want to optimise as the rotation
angles $A$ and $B$ of the second and third gates as $\mathcal{R}_2:= \mathcal{R}(A)$ and $\mathcal{R}_3:= \mathcal{R}(B)$
whereby the angles satisfy $0 < \theta < A < B < 2\pi$ by definition.
The relevant quasiprobabilities $\underline{\gamma}$ satisfy the equation
\begin{equation}\label{eq:quasiprob_general}
	\mathcal{R}(\theta) = \gamma_1(\theta) \mathcal{R}(0) + \gamma_2(\theta) \mathcal{R}(A)
	+  \gamma_3(\theta) \mathcal{R}(B).
\end{equation}
This system of equations is solved by the coefficients
\begin{align*}
\gamma_1=	&\csc \left(\frac{A}{2}\right) \csc \left(\frac{B}{2}\right) \sin \left(\frac{A}{2}{-}\frac{\theta }{2}\right) \sin \left(\frac{B}{2}{-}\frac{\theta }{2}\right),\\
\gamma_2=	&\csc \left(\frac{A}{2}\right) \sin \left(\frac{\theta }{2}\right) \csc \left(\frac{B}{2} {-} \frac{A}{2}\right) \sin \left(\frac{B}{2}{-}\frac{\theta }{2}\right),	\\
	\gamma_3=&- \sin \left(\frac{\theta }{2}\right) \csc \left(\frac{B}{2}\right) \sin \left(\frac{A}{2}{-}\frac{\theta }{2}\right) \csc \left(\frac{B}{2} {-} \frac{A}{2}\right).
\end{align*}
The above solution allows us to analytically calculate the $\ell^1$ norm of the coefficient vector as
\begin{align*}
	\lVert \gamma \rVert_1 &= \frac{1}{2} \csc \left(\frac{B}{2}\right) \csc \left(\frac{A}{2}-\frac{B}{2}\right) \\
	&\times \left(-2 \cos \left(\frac{A}{2}-\theta \right)+\cos \left(\frac{A}{2}-B\right)+\cos \left(\frac{A}{2}\right)\right).
\end{align*}
We now minimise this vector norm as a function of the rotation angles $A$ and $B$

\noindent \textbf{Optimising the rotation angle $A$:}
Let us compute the derivative of the norm $\lVert \gamma \rVert_1$ with respect to $A$ as
\begin{equation*}
 \frac{ \partial \lVert \gamma \rVert_1 } {\partial A} =	\sin \left(\frac{\theta }{2}\right) \csc \left(\frac{B}{2}\right) \csc ^2\left(\frac{A-B}{2}\right) \sin \left(\frac{B-\theta}{2}\right).
\end{equation*}
Indeed, we find that $ \frac{ \partial \lVert \gamma \rVert_1 } {\partial A} > 0$ in the
relevant region where $0 < \theta < A < B$ confirming that the norm increases monotonically
as we increase $A$.
Thus we can monotonically decrease the norm by decreasing $A$ and ultimately approaching the
limit $\lim_{A \rightarrow \theta }\lVert \gamma \rVert_1  = 1$. 
Thus, in order to minimise $\lVert \gamma \rVert_1$ we need to choose $A$ as small as possible
and indeed our best option is choosing the nearest notch setting $A=\Theta_{k+1}$.

\noindent \textbf{Optimising the rotation angle $B$:}
We now find the minimum of the vector norm with respect to the rotation angle $B$.
For this reason we calculate the derivative of the norm with respect to $B$ and solve the trigonometric equation
\begin{align*}
	\frac{ \partial \lVert \gamma \rVert_1 } {\partial B} &=
	\sin \left(\frac{\theta }{2}\right) \csc ^2\left(\frac{B}{2}\right) \sin \left(\frac{A-\theta}{2}\right) \\
	&\times \sin \left(\frac{A}{2}-B\right) \csc ^2\left(\frac{A-B}{2}\right) =0	.
\end{align*}
The above product of trigonometric functions is only $0$ when $ \sin \left(\frac{A}{2}-B\right) =0 $ which equation
is uniquely solved by $B = \pi + A/2$ in the interval $A < B \leq 2\pi$ and indeed corresponds to the
unique minimum along $B$.
\end{proof}

\section{Non-uniform notch settings}

\cref{theo2}
can be directly applied to the scenario when the rotation angles $\{ \Theta_q \}$ are not uniformly distributed
through the general solution of the system of equations in \cref{eq:quasiprob_general}.
In particular, in the main text we assumed that $ \Theta_q$ are equidistantly spaced over the interval $0 \leq  \Theta_q < 2\pi$.
As we detailed in \cref{sec:gen}, in an experiment one might prefer to calibrate $\Theta_q$ to be non-equidistant.
In such a scenario, \cref{theo2} implies that given a desired rotation angle $\theta$, one should choose
the two nearest notch settings $\Theta_{k}$ and $\Theta_{k+1}$ as well as the one nearest to the polar opposite
via the weights $\underline{\gamma}$ computed in \cref{theo2}.

	\section{Symmetries and further solutions \label{app:symm}}
	
	Recall that in \cref{eq:rots} assumed a set of three notch settings is used $\mathcal{R}(\Theta_k), \mathcal{R}(\Theta_{k+1})$ and $\mathcal{R}(\Theta_k {+} \pi)$.
	We can similarly find an optimal solution by assuming the third gate is instead replaced with the rotation angle $\Theta_{k+1} {+} \pi$
	according to \cref{theo1}.
	In this case we can solve for the coefficients and find a different expression compared to \cref{eq:solution} as
	\begin{align*}
		\gamma_1'(\theta) &= \csc (\Delta) \sin (\Delta-\theta )\\
		\gamma_2'(\theta) &= \csc \left(\frac{\Delta}{2}\right) \sin \left(\frac{\theta }{2}\right) \cos \left(\frac{\Delta}{2}-\frac{\theta }{2}\right)\\
		\gamma_3'(\theta) &= -\sec \left(\frac{\Delta}{2}\right) \sin \left(\frac{\theta }{2}\right) \sin \left(\frac{\Delta}{2}-\frac{\theta }{2}\right).\\
	\end{align*}
	However, one can straightforwardly show that the above solution is actually equivalent to \cref{eq:solution}
	up to the symmetry transformation $\theta \rightarrow \Delta - \theta$ and relabelling of the gates,
	i.e., $\gamma_1'(\Delta - \theta) = \gamma_2(\theta)$ and $\gamma_2'(\Delta - \theta) = \gamma_1(\theta)$
	while $\gamma_3' = \gamma_3$. This indeed confirms that we are free to choose
	whether we define $\theta$ as an overrotation of one
	of the two nearest notch settings or as an underrotation of the other nearest notch setting.

	Furthermore,  both $\vec{\gamma}$ and $\vec{\gamma}'$ are solutions to our problem with identical $\ell^1$ norm;
	Since linear combinations of solutions are similarly solutions, there are infinitely
	many solutions that use four rotation angles $\Theta_k$, $\Theta_{k+1}$, $\Theta_{k}+\pi$ and $\Theta_{k+1}+\pi$
	however, it is straightforward to check that any admissible linear combination of these solutions
	have the same $\ell^1$ norm -- which is identical to \cref{eq:l1_norm}.
	We have verified these expectations:
	we set up the system of 3 equations in \cref{eq:channel_form} but with four unknowns $\gamma_1, \gamma_2, \gamma_3$
	and $\gamma_4$, and analytically solved for the unknowns.
	Indeed there are infinitely many solutions but the $\ell^1$ norm of all solutions is equivalent to \cref{eq:l1_norm}.
	Thus, in order to reduce engineering complexity, we
	strongly prefer our optimal solution that uses only 3 notch settings.

One could of course use more than $n = 3$ or $n = 4$ notch settings, which again
results in an underdetermined system of equations as
\cref{eq:channel_form} would contain only three equations for the three degrees of freedom, namely, $1\pm\cos(\theta)$ and $\sin(\theta)$,
while the system would contain $n$ unknowns, namely $\gamma_1,  \dots \gamma_n$.
Our best strategy is then to choose solutions that have minimal $\ell^1$ norms. In fact it is a well-established principle in the
theory of underdetermined systems that the solutions with minimal $\ell^1$ norm are almost always the sparsest possible solutions.
We confirmed numerically that indeed
the solutions with least $\ell^1$ norm are always 3-sparse and correspond to the 3 rotation gates that we chose
in the present work.

\section{Combining with quantum error mitigation \label{app:qem}}

	Error mitigation techniques are straightforwardly compatible
with PAI. Given these techniques are completely decoupled from 
the present approach, error mitigation can be implemented seamlessly 
"on top" of PAI. As error mitigation yields measurement overheads 
that generally grow exponentially with both the number $\nu$ of gates and with the
per-gate error rates $\epsilon$, we detail that the discretisation $\Delta$
should be engineered in accordance with the gate errors $\epsilon$.

\subsection{Common error mitigation techniques}
	In the present work we focused on ideal, unitary gates $\mathcal{R}_1$, $\mathcal{R}_2$ and $\mathcal{R}_3$
	that are used to implement an arbitrary rotation angle $\mathcal{R}(\theta)$ in expectation via \cref{eq:rots}.
	In a real experiment, however, these gates are noisy which we reflect via the tilde notation $\tilde{\mathcal{R}}_j$,
	and as we now detail, nearly all existing
	error mitigation techniques can straightforwardly be combined with the present approach.
		
	First, one can use efficient schemes, such as sparse Pauli--Lindblad or learning-based QEM techniques~\cite{berg2022probabilistic,kim2023evidence,cai2022quantum}
	to efficiently learn the noise model of the local gates $\tilde{\mathcal{R}}_j$;
	one then implements standard Probabilistic Error Cancellation (PEC) techniques~\cite{gambetta_error_mitig,practical_QEM, cai2022quantum}
	to effectively obtain the noise-free unitary gate via the quasiprobability decomposition
	$\mathcal{R}_j = \sum_q \Gamma_{jq} \tilde{U}_q$. Here $\tilde{U}_q$ are native noisy gate operations supported
	by the quantum hardware, e.g., noisy rotations $\tilde{\mathcal{R}}_j$ followed by recovery operations~\cite{cai2022quantum}.
	This PEC approach introduces a measurement overhead due to the increase of variance of observables
	and thus
	 each noisy gate variant $\tilde{\mathcal{R}}_j$ has a measurement overhead $\lVert \Gamma_{j} \rVert_1$
	associated with the cost of mitigating incoherent errors.
	Thus, the measurement overhead $\lVert \gamma(\theta) \rVert_1^2$ to implement the desired noise-free and continuous angle rotation 
	$\mathcal{R}(\theta)$ 
	associated with PAI
	is increased in the worst case to the product
	$\lVert \gamma(\theta) \rVert_1 \, \lVert \gamma_{max} \rVert_1$.
	-- and we will denote as $\lVert \Gamma_{max} \rVert_1 = \max_j \lVert \Gamma_{j} \rVert_1$
	the largest overhead due to gate noise.

	Another commonly used error mitigation technique is zero-noise extrapolation~\cite{cai2022quantum}.
	In this approach one increases the noise level $\epsilon$ of the gates $\tilde{\mathcal{R}}_j(\epsilon)$
	(for example via the previously learned noise models~\cite{kim2023evidence})
	to measure expectation values at increasing error rates; One then extrapolates the expected values to zero error
	to effectively obtain the expectation value one would measure having access to the noise-free gates $\tilde{\mathcal{R}}_j(\epsilon \rightarrow 0)$.
	Finally, purification-based techniques can also be applied straightforwardly~\cite{PhysRevX.11.031057, cai2022quantum}:
	these techniques are oblivious to the error models of the gates
	and by preparing $n$ copies of the noisy quantum state one can guarantee
	that on average the deviation from the desired ideal
	gates $\mathcal{R}_j$ is effectively exponentially small in $n$.
	
\subsection{Measurement overhead}	
	The measurement overhead of the above error-mitigation techniques generally scale exponentially
	with the circuit error rate $\xi = \nu \epsilon$ which is the per-gate error rate $\epsilon$ multiplied by the number $\nu$ of gates. 
	For example, the measurement overhead of PEC grows as
	$e^{4 \nu \epsilon}$ as derived in~\cite{cai2022quantum} for a common error model.
	As we proved in \cref{app:variance}, the overhead associated with PAI
	grows as $e^{ \nu \Delta_{max}^2/4 } $
	and thus the total measurement overhead grows as the product
	\begin{equation*}
		e^{4 \nu \epsilon}
		e^{ \nu \Delta_{max}^2/4 } 
		=
		e^{\nu (4 \epsilon + \Delta_{max}^2/4)}.
	\end{equation*}
	The above dependence on the number $\nu$ of gates
	motivates us to choose a discretisation $\Delta_{max}$
	that ensures the condition $\Delta_{max}^2/4 \approx 4 \epsilon$.
	In particular, choosing too many digits results in a fine discretisation as $\Delta_{max}^2/4 \ll 4 \epsilon$
	which 
	risks overengineering the quantum device, i.e., one could simply increase $\Delta_{max}$
	without significantly increasing the measurement overhead which is dominated by the overhead
	associated with error mitigation.
	In contrast choosing too few digits $\Delta_{max}^2/4 \gg 4 \epsilon$ leads to the measurement
	overhead being dominated by the implementation of PAI.

	Specifically, first generations of devices with error rates $10^{-3} \leq \epsilon \leq 10^{-2}$
	can be engineered in principle with as low as $B=5-6$ digits of precision.
	In contrast, achieving practical quantum advantage will likely require per-gate error rates
	$10^{-4} \leq \epsilon \leq 10^{-3}$
	for which $B=6$ bits of precision may already be sufficient while $B=7$ (and $\epsilon \approx 10^{-4}$)
	would enable useful applications of  early quantum computers.


%

\end{document}